\documentstyle[aps,preprint,amssymb,epsfig]{revtex}

\begin{document}

\draft
\tightenlines
\makebox[\textwidth][r]{SNUTP 97-053}

\begin{center}
{\large\bf
Magnetic Catalysis in Quantum Electrodynamics} 
\vskip 0.2in
Deog Ki Hong\footnote{Email address:$\>$\tt dkhong@hyowon.cc.pusan.ac.kr}\\
\vskip 0.2in 
\it 
Department of Physics, Pusan National
University \protect\\
Pusan 609-735, Korea
\end{center}

\begin{abstract}
We derive the (Wilsonian) low energy effective Lagrangian for
Quantum Electrodynamics under external constant magnetic field by
integrating out all electrons except those in the lowest Landau
level. We find the one-loop effective Lagrangian contains a
marginal four-Fermi interaction with anomalous
dimension, ${(\ln2)^2\over 2\pi^2}e^4$. Renormalization
group analysis shows that the four-Fermi interaction will break chiral
symmetry in QED if the external magnetic field is
extremely strong, $B>10^{42}$ gauss, or if the Landau gap,
$\sqrt{|eB|}>6.5 \times 10^{10}$ GeV.
\end{abstract}


Recently, it has been shown that external (constant) magnetic field
acts as catalysis of dynamical symmetry breaking in quantum
electrodynamics with or without a non-renormalizable four-Fermi
interaction \cite{gusynin,leung,my}. It is then generalized to
QCD under external chromo-magnetic field \cite{ebert}. The essense of
this magnetic catalysis is that electrons of energy much less than the
Landau gap ($E\ll\sqrt{|eB|}$) are effectively $1+1$
dimensional, since the quantum fluctuations perpendicular to the
external magnetic field are suppressed by ${E\over\sqrt{|eB|}}$. Due
to this dimensional reduction, the critical coupling for dynamical
symmetry breaking becomes zero. Namely, dynamical symmetry breaking
occurs for any arbitrarily weak attraction. This has been shown either
by calculating the vacuum energy or by solving the Schwinger-Dyson
equations for the fermion two-point function. 

On the other hand, dynamical symmetry breaking is believed to occur
when particles interact strongly as in QCD. Indeed, it is found that
four-Fermi interaction of electrons in the lowest Landau level
(LLL) are marginal and the $\beta$-function of four-Fermi
coupling is negative for attractive interaction, leading to strong
attraction at low energy \cite{my}. Thus, the result of Schwinger-Dyson
analysis can be understood in terms of renormalization group (RG). But,
it was unclear how the weak electromagnetic interaction
leads to dynamical symmetry breaking when the four-Fermi interaction
is absent, as shown in the Schwinger-Dyson 
analysis \cite{gusynin,leung,my},
since the Coulomb interaction of electrons remains weak at low  
energy \cite{my}.  
In this paper, we attempt to understand the dynamical
symmetry breaking in pure QED under external magnetic field in terms
of RG analysis.   In this attempt, we find
that, if one integrates out electrons in the higher Landau levels,
there will be a new low-energy effective operator for electron-photon
coupling at tree level, among others, and this operator will generate a
four-Fermi interation at one-loop, which becomes strong in infrared
region.

Magnetic catalysis is a very interesting phenomenon and has 
potential applications in astrophysics such as the cooling process of
neutron stars or particle interactions in early universe under
premodial magnetic field \cite{ns}. 
In order for the magnetic catalysis
to operate, the external magnetic field has to be strong enough
so that the average energy of charged particles is much
smaller than the Landau gap. Namely, the gap has to be
bigger than the rest mass energy, $\sqrt{|eB|}\gg m$, or, 
in the early universe, the temperature if the particles are
relativistic, $\sqrt{|eB|}\gg T$. Therefore the electrons will be
catalyzed only when the external magnetic fields are stronger than a
critical field, $B>B_c$,  where $B_c=m_e^2/|e|\simeq10^{14}$ gauss (G).
If electrons are massless, $B_c=0$ and therefore for any 
weak external magnetic field they will be catalyzed to get a dynamical
mass, $m_{dyn}\simeq \sqrt{|eB|}e^{-1/g}$, where $g$ is the coupling at
the cut-off scale, $\sqrt{|eB|}$. But, the effect is relevant only at
distance larger than $1/m_{dyn}$, which is enormously large for weak
field. In this paper, we consider electrons of energy 
larger than  the rest mass energy but smaller than the Landau gap,
$m<E<\sqrt{|eB|}$, and neglect the electron mass. 

As derived by Schwinger \cite{schwinger}, the electron propagator in a
constant external magnetic field is given as 
\begin{equation}
S(x,y)={\tilde
S(x-y)}\exp\left[{ie\over2}(x-y)^{\mu}A^{\rm ext}_{\mu}(x+y)\right],
\end{equation}
with the Fourier transform of $\tilde{S}$,
\begin{equation}
{\tilde S(k)}=ie^{-k_{\perp}^2/|eB|}\sum_{n=0}^{\infty}(-1)^n
{D_n(eB, k)\over k_{\parallel}^2-2|eB|n},
\end{equation}
where $k_{\perp}$ is the 3-momentum perpendicular to the direction of
the external magnetic field, $k_{\parallel}=k-k_{\perp}$, and  
\begin{equation}
D_n(eB,k)=2\mathord{\not\mathrel{k_{\parallel}}}\left[
P_-L_n\left({2k_{\perp}^2\over
|eB|}\right)-P_+L_{n-1}\left({2k_{\perp}^2\over |eB|}\right) \right]
+4 \mathord{\not\mathrel{k_{\perp}}}
 L_{n-1}^1\left({2k_{\perp}^2\over |eB|}\right).
\end{equation}
$L_n^{\alpha}$ are the associate Laguerre polynomials and $P_{+}$ 
($P_-$) 
is the projection operator which projects out the electrons of
spin (anti-) parallel to the magnetic field direction. 
For $\vec B=B\hat z$, 
$2P_{\pm}=1\pm i\gamma^1\gamma^2 {\rm sign} (eB)$. 

To find the Landau level, we need to solve
the following eigenvalue equation, 
\begin{equation}
[ \vec{\alpha}\cdot({\vec p}- e {\vec A^{\rm ext}}) ]\Psi=E\Psi.
\end{equation}
We take $\vec A^{\rm ext}=(-{B\over2}y,{B\over2}x,0)$. 
The eigenvalues are indexed by collective index $A=(\alpha,\beta,n,k_z)$
and given by 
\begin{eqnarray}
E_A=\alpha\sqrt{k_z^2+ 2\left|eB\right|n}
\end{eqnarray}
where $\alpha=\pm$ denotes the sign of the energy, $\beta=\pm{1\over2}$
is the spin component along the magnetic field, 
and the quantum number $n$ 
is given by 
\begin{eqnarray}
2n=2n_r+1+|m_L|-{\rm sign}(eB) (m_L+2\beta). 
\end{eqnarray}
$n$ is a nonnegative integer that labels the Landau level.
Here $n_r$ is the number of nodes of radial eigenfuction, $m_L$ is the 
angular momentum of the eigenfunction.
The eigenfunction is 
\begin{eqnarray} 
U_A=N_A e^{ik_z
z}e^{im_L\phi}{r_{\perp}}^{|m_L|}L_{n_r}^{|m_L|}(|eB|{r_{\perp}}^2/2)
\exp{(-|eB|{r_{\perp}}^2/4)} u_{\alpha,\beta}, 
\end{eqnarray}
where $N_A $ is the nomalization and the
radial distance ${r_{\perp}}=\sqrt{x^2+y^2}$. 
$L_n^m(x) $ is the associated
Laguerre polynomial  and
$u_{\alpha,\beta}=\chi_\alpha\otimes\eta_\beta$  is the eigenvector of
$\sigma_3\otimes\sigma_3$ where two $\sigma_3$ correspond to the
energy and the spin, respectively. 

Since the eigenfunctions form an orthonormal basis, one can expand the
electron field as 
\begin{equation} 
\Psi(x)=\int\!\!\sum_A \psi_A(t)U_A(\vec x).
\label{mode}
\end{equation}
Now, we divide the electron field into two groups
\begin{equation}
\Psi=\psi+\Psi_{n\ne0},
\end{equation}
where $\psi$ contains only the lowest Landau level ($n=0$ ) in
the mode expansion in Eq.\ (\ref{mode}) and $\Psi_{n\ne0}$ contains the
rest. 

The low energy effective Lagrangian is obtained by integrating out
the higher modes,
\begin{equation}
\exp\left(i\int d^4x{\cal L}_{eff}(\psi,A_{\mu})\right)=\int
{\cal D}\Psi_{n\ne0}\exp\left(i \int d^4x
\left[-{1\over4}F_{\mu\nu}^2+\bar\Psi i
\mathord{\not\mathrel{D}}\Psi \right]\right),
\end{equation}
where the covariant derivative
$D_{\mu}=\partial_{\mu}+ieA_{\mu}+ieA_{\mu}^{\rm ext}$.  
In practice, one derives the
effective Lagrangian by matching all the ``one light particle
irreducible" (1LPI) diagrams systematically in a loop expansion.

\begin{figure}
\vspace{0.5cm}
\centerline{
\epsfig{figure=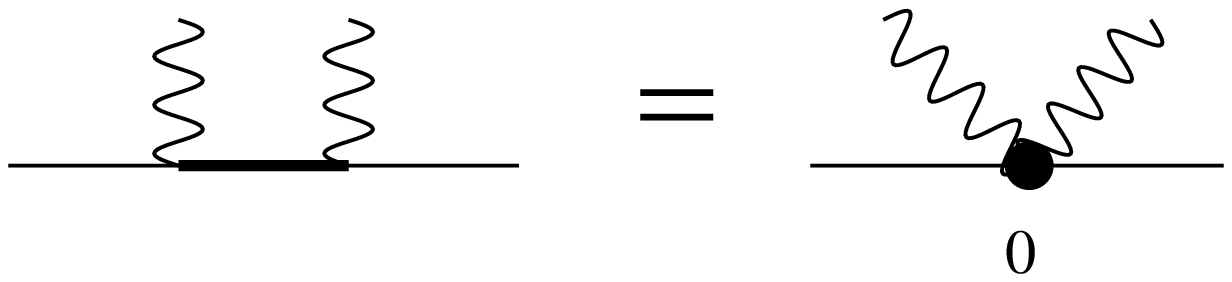,height=2cm}}
\vspace{0.5cm}
\caption{Tree level matching condition.
Diagram on the left is in the full theory,
while the one on the right is
in the effective theory. The heavy line corresponds to electron
in the higher Landau levels, light lines LLL electrons, and
wiggly lines photons; number beneath the vertex counts the loop
order.
}
\label{fig1}
\end{figure}

At tree level, $\Psi_{n\ne0}$ exchange generates a new
photon-electron vertex in the effective theory 
(Fig.\ \ref{fig1});
\begin{eqnarray}
\Gamma_4^0(p_1,p_2,p_1^{\prime},p_2^{\prime}) & 
= & ie^2\bar\psi(p_1)\not\!\! A(p_2)
e^{-k_{\perp}^2/|eB|}\sum_{n=1}^{\infty}(-1)^n {D_n(eB,k)\over
k_{\parallel}^2-2|eB|n}\not\!\!
A(p_1^{\prime})\psi(p_2^{\prime})\\ 
& = & i{e^2\over
|eB|}\bar\psi\not\!\! A\Big[\not\!\! k_{\parallel}
i\gamma_1\gamma_2 \ln2~ {\rm sign}(eB)+\not\!\! k_{\perp} \Big]
\not\! \!A\psi+O\left({k^2\over |eB|}\right),
\end{eqnarray}
where $k=p_1+p_2=p_1^{\prime}+p_2^{\prime}$. In the second line,
we have expanded the electron propagator in powers of momentum
and used $L_{n-1}(0)=1, L_{n-1}^1(0)=n$. The tree-level effective 
Lagrangian is then 
\begin{equation}
{\cal L}_{eff}^0=-{1\over4}F_{\mu\nu}^2+\bar\psi 
(i\mathord{\not\mathrel{D}}
)\psi-{e^2\over 2|eB|}
\bar\psi\gamma^{\alpha}A_{\alpha}{\tilde\gamma}^{\mu}i
\stackrel{\leftrightarrow}\partial_{\mu}
\gamma^{\beta}\psi A_{\beta}+\cdots,
\end{equation}
The ellipsis refers to operators containing more photons and
higher power of derivatives, and 
\begin{equation}
\tilde\gamma^{\mu}=
    \left\{ \begin{array}{ll}
	i\gamma^{\mu}\gamma^1\gamma^2\ln2~ {\rm sign}(eB) &
        \mbox{if $\mu=0,3$}\\
        \gamma^{\mu} & \mbox{if $\mu=1,2$}.
    \end{array}
    \right.
\end{equation} 

\begin{figure}
\vspace{0.5cm}
\centerline{
\epsfig{figure=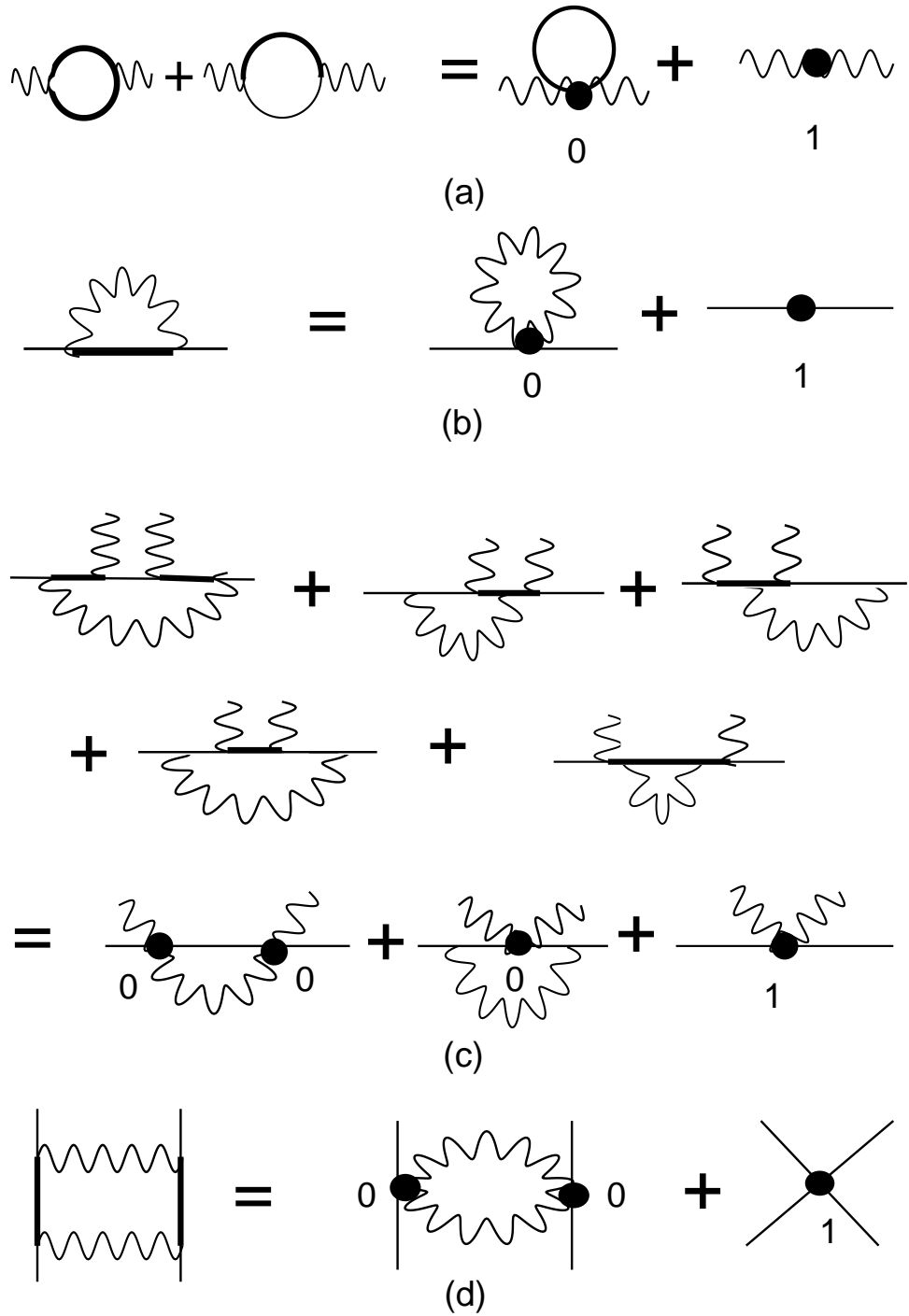,height=20cm}}
\vspace{0.5cm}
\caption{One-loop matching conditions}
\label{fig2}
\end{figure}

To match at one-loop (see Fig.\ \ref{fig2}), 
we have to consider graphs with
two, three, four, and more external fields to get 
\begin{eqnarray}
{\cal L}_{eff}^1 & = &-{1\over4}(1+a_1)
F_{\mu\nu}^2+(1+b_1)\bar\psi (i\mathord{\not\mathrel{D}}
)\psi
-(1+c_1){ie^2\over |eB|}\bar\psi\mathord{\not\mathrel{A}}
{\tilde\gamma}_{\mu}
\stackrel{\leftrightarrow}\partial^{\mu}_{\parallel}
\mathord{\not\mathrel{A}}\psi \nonumber \\ 
&  & -{ie^2\over |eB|}
\bar\psi\mathord{\not\mathrel{A}} {\tilde\gamma}_{\mu}
\stackrel{\leftrightarrow}\partial^{\mu}_{\perp}
\mathord{\not\mathrel{A}}\psi 
+{g_1\over|eB|}\left[\left(\bar\psi\psi\right)^2+
\left(\bar\psi i\gamma_5\psi\right)^2\right]+\cdots,
\end{eqnarray}
where the coefficients $a_1, b_1, c_1, g_1$ are
dimensionless and of order $e^2$. 
The
ellipsis denotes terms with more external fields and derivatives.
Note that the new electron-photon effective couplings get different
corrections, depending on the component of the total momentum carried
by electron and photon. 

We find that 
the four-Fermi operator arises, in a chirally invariant
form, at one-loop when we match four-Fermion amplitudes both in the
full theory and  in the effective theory. 
The one-loop amplitude for four-Fermi interaction in the full theory 
is 
\begin{equation}
{\cal A}_4=i{3fe^4\over 32\pi^2|eB|}
\left[\left(\bar\psi\psi\right)^2+
\left(\bar\psi i\gamma_5\psi\right)^2\right].
\end{equation}
The amplitude should converge since the Feynman diagram is finite
and, in general, the dimesionless number $f$ will be of order one.

In the effective theory,  the one-loop amplitude is                   
\begin{equation}
i{g_1\over 4|eB|}
\left[\left(\bar\psi\psi\right)^2+
\left(\bar\psi i\gamma_5\psi\right)^2\right],
\end{equation}
with 
\begin{equation}
g_1={e^4(\ln2)^2\over 4\pi^2}
\left[-{1\over\epsilon}-1.34+\gamma-{\pi^2\over8}+\ln{|eB|\over
\mu^2} \right]+\delta g_1,
\end{equation}
where $\delta g_1$ is the one-loop counter term
for the four-Fermi amplitude to remove the pole divergence and to
ensure the matching at the cutoff scale, $\mu=\sqrt{|eB|}$, which is 
\begin{equation}
g_1(\mu=\sqrt{|eB|}-0)={3fe^4\over 8\pi^2} 
\end{equation}

After matching the operators in the effective
theory with those in the full theory at $\sqrt{|eB|}$, we run the
operators down to the scale of interest. We find that $a_1,b_1$ do not
scale as we expect from the Ward-Takahashi identity because the
electric charge does not run at low energy due to the dimensional
reduction. But, $c(\mu)=1+c_1+\cdots$ and $g(\mu)=g_1+\cdots$ do. 
To get the right value for the strength of the four-Fermi operator, 
we have to rescale the fields to get a conventionally normalized 
kinetic term,
\begin{equation}
\psi(x)\longrightarrow \psi^{\prime}(x)=\sqrt{1+b_1}\psi(x),
\end{equation}                                              
and take into account of the scale dependence of $c_1(\mu)$. 
But, since these are higher-order corrections to $g_1$, the
coefficient of the four-Fermi operator at $\mu=\sqrt{|eB|}$ is 
just ${3fe^4\over8\pi^2}$ in the leading order. 
As shown in
\cite{my}, this four-Fermi operator is marginal: 
Under the scale transformation, $E\to sE$, the
electron should transform as $\psi\to s^{1/2}\psi$, since  the radial
dependence of $\psi$ is fixed as $\exp(-{r_{\perp}}^2|eB|/4)$ and
therefore the coordinate $x,y$ should not scale. The
dimensional parameter, $|eB|$, in front of the operator will cancel
out upon the momentum integration in perturbation theory, leaving
the dimensionless parameter $g=g_1+O(\hbar^2)$ as the expansion
parameter. 

Since the energy dispersion of photon is 
$E=\pm\sqrt{k_z^2+k_{\perp}^2}$, photon with momentum parallel to the
magnetic field ($k_{\perp}=0$) has the scaling dimension $0$ 
under the scale transformation, $k_{\parallel}\to s
k_{\parallel}$ and $k_{\perp}\to k_{\perp}$, 
while photon with nonvanishing $k_{\perp}$ has the scaling dimension
1. Therefore, the new photon-electron coupling is marginal if the
photon carries no perpendicular momentum component. But, it turns
out that its anomalous dimension is negative, $\mu{\partial
\over\partial\mu}c_1(\mu)=e^2\ln2/\pi^2>0$. Therefore, 
the coupling becomes irrelevant if we include the one-loop
correction. On the other hand, the
four-Fermi interaction becomes relevant at quantum level.
Below the cutoff scale, the renormalization group equation for $g_1$, 
\begin{equation}
\mu{d\over d\mu}g_1=-{(\ln2)^2\over 2\pi^2}e^4,
\end{equation}
which is very small. 
Hence, $g_1$ remains almost constant as the scale changes. 
Integrating this equation from $\sqrt{|eB|}$ to $m_e$ gives 
\begin{equation}
g_1(m_e)= g_1(\sqrt{|eB|})+4\alpha^2(\ln2)^2
\ln {|eB|\over m_e^2}, 
\label{four}
\end{equation}
where $\alpha=e^2/4\pi$ is the fine-structure constant
at scale $\sqrt{|eB|}$.

In this picture, 
the QED coupling runs down till the Landau gap scale. Below the
scale it does not run, but a new marginal
four-Fermi coupling appears due to quantum effects of the higher
level modes and gets enhanced at low energy. 
In order for the four-Fermi coupling gets strong enough to
break chiral symmetry, 
the coupling has to run extremely long and become 
of order 1 at $\mu\simeq m_e$. 
To find the lower bound for the Landau gap that allows 
sufficient evolution for the four-Fermi interaction to break   
chiral symmetry, we need to know the value of the fine-structure 
constant, $\alpha$, at the Landau gap scale, which is given 
for $\sqrt{|eB|}>M_Z$ as \cite{georgi,ramond}   
\begin{equation}
{1\over \alpha(\sqrt{|eB|})}={1\over\alpha(M_Z)}-
{1\over2\pi}\cdot{14\over 15}\ln\left({\sqrt{|eB|}\over M_Z}\right), 
\label{qed}
\end{equation}
where $\alpha(M_Z)=0.00782$, 
the $Z$ mass, $M_Z=91.175\pm0.021$ GeV and we assume 
the standard model particle content.  
Combining Eq.'s\ (\ref{four}) and (\ref{qed}), we find that 
$\alpha(\sqrt{|eB|})\simeq1/124.8$ and the low energy effective 
four-Fermi coupling becomes of order one 
if $|eB|>1.6 \times 10^{28}~m_e^2\simeq 4.2 \times 10^{21}$
${\rm GeV}^2$, or $B>10^{42}$ G. 
(Note that in the effective theory the value of electric charge 
is given by the value at $\mu=\sqrt{|eB|}$.)

Therefore, the effective four-Fermi
interaction will break the chiral symmetry of QED 
if $B>10^{42}$ G, 
which is much stronger than any known magnetic
field in nature.   The dynamical mass will be then 
\begin{equation}
m_{\rm dyn}\simeq \sqrt{|eB|}\exp\left(-{2\over (\ln2)^2
\alpha^2}\right).
\end{equation}
Note that this result is different from the one obtained from
Schwinger-Dyson analysis, $m_{\rm dyn}\simeq
\sqrt{|eB|}\exp\left(-{\pi^2\over\sqrt{2}|e|}\right)$ \cite{gusynin}.
This is expected, since, according to RG analysis, the relevant
operator  needed for infrared instability occurs by quantum
effects of electrons in the higher Landau levels, which is higher 
order in coupling constant expansion, while the Schwinger-Dyson
analysis shows dynamical mass is generated soley
by the electrons in the lowest Landau level.
But, in the Schwinger-Dyson analysis, one has to integrate 
the loop momentum over all ranges. Hence, the effective four-Fermi 
interaction may not be negligible if one includes the running 
effect of the coupling, going beyond the ladder approximation. 
Therefore, it will be interesting to solve the Schwinger-Dyson 
equations in the effective Lagrangian derived in this paper.

In conclusion, we derive the (one-loop) 
low energy effective Lagrangian
for Quantum Electrodynamics under a constant external magnetic field
by integrating out the higher Landau level modes ($n\ne0$). We find
that the effective Lagrangian includes a new electron-photon coupling
which  generates a marginal four-Fermi interaction  at one-loop
level.  Since the anomalous dimension of the four-Fermi interaction is
positive though very small, about $2.2\times10^{-4}$, it gets
enhanced at low energy.  We find that the effective four-Fermi
interaction will break chiral symmetry if the external magnetic
field is extremely strong, $B>1.6\times10^{28}{m_e^2\over|e|}\simeq
10^{42}$ G. 

\acknowledgments

We thank K. Choi and V. Gusynin for discussions and H.C. Kim  
for the help in drawing figures. 
This work was supported in part by the KOSEF through SRC program of 
SNU-CTP and also by Basic Science Research Program, Ministry of 
Education, 1996 (BSRI-96-2413).

\end{document}